\documentclass[aps, prc,twocolumn,superscriptaddress,groupedaddress]{revtex4}  
\usepackage{graphicx}  
\usepackage{dcolumn}   
\usepackage{bm}        
\usepackage{amssymb}   
\usepackage{siunitx}   
\usepackage{natbib}
\usepackage{multirow}
\usepackage{xcolor}
\usepackage{braket}
\usepackage{booktabs}
\usepackage{subfigure}

\usepackage[english]{babel}
\usepackage[T1]{fontenc}
\usepackage[utf8]{inputenc}

\usepackage[colorlinks,hyperindex]{hyperref}

\hypersetup{colorlinks=true, linkcolor=blue, citecolor=blue}

\begin{document}



\title{\textsuperscript{20}Ne~+~\textsuperscript{76}Ge elastic and inelastic scattering at 306~MeV}





\author{A.~Spatafora}
\email{alessandro.spatafora@lns.infn.it}
\affiliation{Istituto Nazionale di Fisica Nucleare,  Laboratori Nazionali del Sud, Catania, Italy}
\affiliation{Dipartimento di Fisica e Astronomia "Ettore Majorana", Universit\`a di Catania, Catania, Italy}

\author{F.~Cappuzzello}
\email{cappuzzello@lns.infn.it}
\affiliation{Istituto Nazionale di Fisica Nucleare,  Laboratori Nazionali del Sud, Catania, Italy}
\affiliation{Dipartimento di Fisica e Astronomia "Ettore Majorana", Universit\`a di Catania, Catania, Italy}

\author{D.~Carbone}
\affiliation{Istituto Nazionale di Fisica Nucleare,  Laboratori Nazionali del Sud, Catania, Italy}

\author{M.~Cavallaro}
\affiliation{Istituto Nazionale di Fisica Nucleare,  Laboratori Nazionali del Sud, Catania, Italy}

\author{J.~A.~Lay}
\affiliation{Departamento de F\'isica At\'omica Molecular y Nuclear, Universidad de Sevilla, Sevilla, Spain}
\affiliation{Instituto Interuniversitario Carlos I de F\'isica Te\'orica y Computacional (iC1), Sevilla, Spain}

\author{L.~Acosta}
\affiliation{Instituto de Física, Universidad Nacional Aut\'onoma de M\'exico, 04510  M\'exico City, M\'exico.}

\author{C.~Agodi}
\affiliation{Istituto Nazionale di Fisica Nucleare,  Laboratori Nazionali del Sud, Catania, Italy}

\author{D.~Bonanno}
\affiliation{Istituto Nazionale di Fisica Nucleare,  Sezione di Catania, Catania, Italy}

\author{D.~Bongiovanni}
\affiliation{Istituto Nazionale di Fisica Nucleare,  Laboratori Nazionali del Sud, Catania, Italy}

\author{I.~Boztosun}
\affiliation{Akdeniz University, Antalya, Turkey}

\author{G.~A.~Brischetto}
\affiliation{Istituto Nazionale di Fisica Nucleare,  Laboratori Nazionali del Sud, Catania, Italy}
\affiliation{Dipartimento di Fisica e Astronomia "Ettore Majorana", Universit\`a di Catania, Catania, Italy}
\affiliation{Centro Siciliano di Fisica Nucleare e Struttura della Materia, Catania, Italy}

\author{S.~Burrello}
\affiliation{Istituto Nazionale di Fisica Nucleare,  Laboratori Nazionali del Sud, Catania, Italy}

\author{S.~Calabrese}
\affiliation{Istituto Nazionale di Fisica Nucleare,  Laboratori Nazionali del Sud, Catania, Italy}
\affiliation{Dipartimento di Fisica e Astronomia "Ettore Majorana", Universit\`a di Catania, Catania, Italy}

\author{D.~Calvo}
\affiliation{Istituto Nazionale di Fisica Nucleare,  Sezione di Torino, Torino, Italy}

\author{E.~R.~Ch\`avez Lomel\'i}
\affiliation{Instituto de Física, Universidad Nacional Aut\'onoma de M\'exico, 04510  M\'exico City, M\'exico}

\author{I.~Ciraldo}
\affiliation{Istituto Nazionale di Fisica Nucleare,  Laboratori Nazionali del Sud, Catania, Italy}
\affiliation{Dipartimento di Fisica e Astronomia "Ettore Majorana", Universit\`a di Catania, Catania, Italy}

\author{M.~Colonna}
\affiliation{Istituto Nazionale di Fisica Nucleare,  Laboratori Nazionali del Sud, Catania, Italy}

\author{F.~Delaunay}
\affiliation{Laboratoire de Physique Corpuscolaire de Caen, Normandie Universit\'e, ENSICAEN, UNICAEN, CNRS/IN2P3, Caen France}
\affiliation{Istituto Nazionale di Fisica Nucleare,  Sezione di Torino, Torino, Italy}
\affiliation{Dipartimento Scienza Applicata e Tecnologia, Politecnico di Torino, Torino, Italy}

\author{N.~Deshmukh}
\affiliation{Istituto Nazionale di Fisica Nucleare,  Laboratori Nazionali del Sud, Catania, Italy}

\author{J.~L.~Ferreira}
\affiliation{Instituto de Fisica, Universidade Federal Fluminense, Niteroi, Brazil}

\author{P.~Finocchiaro}
\affiliation{Istituto Nazionale di Fisica Nucleare,  Laboratori Nazionali del Sud, Catania, Italy}

\author{M.~Fisichella}
\affiliation{Istituto Nazionale di Fisica Nucleare,  Sezione di Torino, Torino, Italy}

\author{A.~Foti}
\affiliation{Istituto Nazionale di Fisica Nucleare,  Sezione di Catania, Catania, Italy}

\author{G.~Gallo}
\affiliation{Istituto Nazionale di Fisica Nucleare,  Laboratori Nazionali del Sud, Catania, Italy}
\affiliation{Dipartimento di Fisica e Astronomia "Ettore Majorana", Universit\`a di Catania, Catania, Italy}

\author{A.~Hacisalihoglu}
\affiliation{Istituto Nazionale di Fisica Nucleare,  Laboratori Nazionali del Sud, Catania, Italy}
\affiliation{Institute of Natural Science, Karadeniz Teknik Universitesi, Trabzon, Turkey}

\author{F.~Iazzi}
\affiliation{Istituto Nazionale di Fisica Nucleare,  Sezione di Torino, Torino, Italy}
\affiliation{Dipartimento Scienza Applicata e Tecnologia, Politecnico di Torino, Torino, Italy}

\author{G.~Lanzalone}
\affiliation{Istituto Nazionale di Fisica Nucleare,  Laboratori Nazionali del Sud, Catania, Italy}
\affiliation{Universit\`a degli Studi di Enna "Kore", Enna, Italy}

\author{H.~Lenske}
\affiliation{University of Giessen, Giessen, Germany}

\author{R.~Linares}
\affiliation{Instituto de Fisica, Universidade Federal Fluminense, Niteroi, Brazil}

\author{D.~Lo Presti}
\affiliation{Dipartimento di Fisica e Astronomia, Universit\`a di Catania, Catania, Italy}
\affiliation{Istituto Nazionale di Fisica Nucleare,  Sezione di Catania, Catania, Italy}

\author{J.~Lubian}
\affiliation{Instituto de Fisica, Universidade Federal Fluminense, Niteroi, Brazil}

\author{M.~Moralles}
\affiliation{Instituto de Pesquisas Energeticas e Nucleares IPEN/CNEN, Sao Paulo, Brazil}

\author{A.~Muoio}
\affiliation{Istituto Nazionale di Fisica Nucleare,  Laboratori Nazionali del Sud, Catania, Italy}

\author{J.~R.~B.~Oliveira}
\affiliation{Instituto de Fisica, Universidade de S\~ao Paulo, S\~ao Paulo, Brazil}

\author{A.~Pakou}
\affiliation{Department of Physics, University of Ioannina and Hellenic Institute of Nuclear Physics, Ioannina, Greece}

\author{L.~Pandola}
\affiliation{Istituto Nazionale di Fisica Nucleare,  Laboratori Nazionali del Sud, Catania, Italy}

\author{H.~Petrascu}
\affiliation{Institutul National de Cercetare-Dezvoltare pentru Fizica si Inginerie Nucleara Horia Hulubei, Bucarest, Romania}

\author{F.~Pinna}
\affiliation{Istituto Nazionale di Fisica Nucleare,  Sezione di Torino, Torino, Italy}
\affiliation{Dipartimento Scienza Applicata e Tecnologia, Politecnico di Torino, Torino, Italy}

\author{S.~Reito}
\affiliation{Istituto Nazionale di Fisica Nucleare,  Sezione di Catania, Catania, Italy}

\author{G.~Russo}
\affiliation{Dipartimento di Fisica e Astronomia "Ettore Majorana", Universit\`a di Catania, Catania, Italy}
\affiliation{Istituto Nazionale di Fisica Nucleare,  Sezione di Catania, Catania, Italy}

\author{G.~Santagati}
\affiliation{Istituto Nazionale di Fisica Nucleare,  Laboratori Nazionali del Sud, Catania, Italy}

\author{O.~Sgouros}
\affiliation{Istituto Nazionale di Fisica Nucleare,  Laboratori Nazionali del Sud, Catania, Italy}

\author{S.~O.~Solakci}
\affiliation{Akdeniz University, Antalya, Turkey}

\author{V.~Soukeras}
\affiliation{Istituto Nazionale di Fisica Nucleare,  Laboratori Nazionali del Sud, Catania, Italy}

\author{G.~Souliotis}
\affiliation{Department of Chemistry, University of Athens and Hellenic Institute of Nuclear Physics, Athens, Greece}

\author{D.~Torresi}
\affiliation{Istituto Nazionale di Fisica Nucleare,  Laboratori Nazionali del Sud, Catania, Italy}

\author{S.~Tudisco}
\affiliation{Istituto Nazionale di Fisica Nucleare,  Laboratori Nazionali del Sud, Catania, Italy}

\author{A.~Yildirim}
\affiliation{Akdeniz University, Antalya, Turkey}

\author{V.~A.~B.~Zagatto}
\affiliation{Instituto de Fisica, Universidade de S\~ao Paulo, S\~ao Paulo, Brazil}

\collaboration{\large for the NUMEN collaboration}

\begin{abstract}
\textbf{Background:} Double Charge Exchange (DCE) nuclear reactions have recently  attracted much interest as tools to provide experimentally driven information  about the \textit{Nuclear Matrix Elements} of interest in the context of \textit{neutrino-less double beta decay}. In this framework a good description of the reaction mechanism and a complete knowledge of the initial and final state interactions is mandatory. Presently, not enough is known about the details of the optical potentials and nuclear response to isospin operators for many of the projectile-target systems proposed for future DCE studies. Among these, \textsuperscript{20}Ne + \textsuperscript{76}Ge DCE reaction is particularly relevant due to its connection with \textsuperscript{76}Ge double beta decay. \textbf{Purpose:} Characterization of the initial state interaction for the \textsuperscript{20}Ne + \textsuperscript{76}Ge reactions at 306 MeV bombarding energy: determination of the optical potential and exploration of the role of the couplings between elastic channel and inelastic transitions to the first low-lying excited states. \textbf{Methods:} Determination of the experimental elastic and inelastic scattering cross section angular distributions. Comparison of the theoretical predictions adopting  different models of optical potentials with the experimental data. Evaluation of the coupling effect through the comparison of the Distorted Wave Born Approximation calculations with the Coupled Channels ones. \textbf{Results:} Optical Models fail in the description of the elastic angular distribution above the grazing angle ($\sim$~\ang{9.4}). A correction in the geometry to effectively account for deformation of the involved nuclear systems improves the agreement up to about \ang{14}. Coupled channels effects are crucial to obtain a good agreement at large angles in the elastic scattering cross section. \textbf{Conclusions:} The analysis of elastic and inelastic scattering data turned out to be a powerful tool to explore the initial and final state interaction in heavy ion nuclear reactions also at high transferred momenta.  


\end{abstract}

\maketitle

\section{\label{introduction}Introduction}

The study of single (SCE) and double charge exchange (DCE) reaction cross sections induced by heavy projectiles, has recently attracted much interest due to the possible connection with double beta decay \cite{delloro2014}. In particular, the NUMEN and NURE projects at the INFN-LNS \cite{numen2018,numen_proc_braz} aims at extracting relevant information regarding nuclear structure of \textit{neutrino-less double beta decay} ($0\nu\beta\beta$) Nuclear Matrix Elements (NME) by measuring cross sections of DCE and SCE. Other complementary studies at RIKEN  \cite{matsubara_riken,kisamori_riken} and RCNP \cite{takahisa_RNCP} have recently focused on DCE reactions not only in relation to $0\nu\beta\beta$ but also to populate exotic structures. New theoretical indications suggest that not only the transition to the ground state of the daughter nucleus in $0\nu\beta\beta$, accessed by DCE reaction, but also the whole double Gamow-Teller strength could, in principle, be connected to $0\nu\beta\beta$-NME \cite{shimizu}. 

This task demands that a detailed microscopic description of the reaction mechanism is accomplished. The state-of-art reaction theory for the description of SCE and DCE reactions up to high excitation energy triggers the need of a careful knowledge of initial and final state interactions associated to the involved beams, e. g. \textsuperscript{12}C, \textsuperscript{18}O and \textsuperscript{20}Ne, at energies from 10 to 60 MeV/u and the heavy targets of interest for the NUMEN purposes. The information about the optical potential for such systems in this energy regime is limited and not deeply tested. This task in turn requires the control of the distortions of the incoming and outgoing waves due to the optical potentials, as shown in refs. \cite{lenske2018,lenske_CNNP, colonna}.

Optical potentials play a central role in heavy-ion quantum scattering theory, as they describe the average interaction during the collisions between the nuclei involved in the entrance and exit partitions. The complete many-nucleon scattering is a challenging problem. Very recently, \textit{ab initio} reaction theory \cite{kumar2017} has shown to be a promising theory but still limited to light nuclear systems with few reaction channels open. The complexity of the many-body scattering problem of heavy nuclei can be carried out to a calculable scheme by the choice of an appropriate model space for the nuclear wave functions, with the caveat that the space left out from the model is effectively accounted by the introduction of a complex polarization potential \cite{satchler}. Such a procedure has been commonly used to analyses of atomic nuclei elastic and inelastic scattering such as reaction channels. For the latter the optical potential is complemented by a perturbative potential, describing the specific nuclear transitions feeding the selected outgoing channel.

A typical way to determine the optical potential is to use parametrizations of Woods-Saxon shape adjusted to reproduce experimental elastic scattering cross sections at the appropriate incident energy and momentum transfer for the nuclear systems involved in both the incoming and outgoing channels. Despite its simplicity, this procedure hides the role of internal degrees of freedom of the projectile and target nuclei.  Potentials obtained by folding the \textit{frozen} densities of the colliding nuclei with a realistic nucleon-nucleon interaction are expected to be successful models of the optical potentials. This is mainly due to the strong absorption which confines the reaction source close to the surfaces of the colliding systems, making the system less sensitive to the internal regions where the overlap between projectile and target can change significantly the densities from the assumed frozen condition \cite{satchler1979}. Although several references \cite{sgouros2013, trz2015}, to our knowledge, no data exist at beam energy of about 15 AMeV and no theoretical studies have been published concerning to the case of \textsuperscript{20}Ne beam, which is one of the main probes for NUMEN. 

A deeper understanding of the scattering dynamics can be achieved using the \textit{coupled channels} (CC) method, in which relevant internal states of the projectile/target systems are explicitly taken into account. An interesting application of the CC analysis is the recent search of nuclear structure effects in \textsuperscript{12}C+\textsuperscript{12}C, \textsuperscript{16}O+\textsuperscript{16}O at energies around the Coulomb barrier up to 1.5 GeV, where the complex diffractive patterns observed in the angular distributions were ascribed to couplings of elastic channel with 2\textsuperscript{+}, 3\textsuperscript{-} and 4\textsuperscript{+} low-lying excited states \cite{ohkubo2014, ohkubo2002}. Similar results are found in \textsuperscript{16}O+\textsuperscript{27}Al and \textsuperscript{16}O+\textsuperscript{60}Ni elastic and inelastic cross sections data  \cite{zagatto2018, pereira_rainbow, oliveira_rainbow, cappuzzello2016}, where the important role of nuclear surfaces deformation and the coupling with inelastic transitions and transfer reactions is also discussed. The main picture is that CC effects are relevant, especially when the interest is in low cross section data, as elastic scattering at large angles or suppressed transitions in the reaction channels. 

This work is part of the network of reactions studied in the same experiment performed at the LNS-INFN within the NUMEN project with the goal to extract the cross-section of the \textsuperscript{76}Ge(\textsuperscript{20}Ne,\textsuperscript{20}O)\textsuperscript{76}Se DCE reaction. Elastic and inelastic scattering data for the system \textsuperscript{20}Ne + \textsuperscript{76}Ge at 15.3 AMeV are here shown. The experimental set-up and the data reduction technique are described in the next sections. The analysis was performed in the Optical Model (OM) framework through the introduction of several optical potentials and in the CC approach. Results obtained with Woods-Saxon like potential following the Akyuz-Winther systematics \cite{akyuzwinther, broglia1991} and double folding potentials are compared to the data. In particular, two different double folding potentials are used: the S\~ao Paulo \cite{chamon2002, pereira2009} potential and a double folding potential using the T-matrix free nucleon-nucleon interaction \cite{lenske_dfol,franey1985}. The effect of lowest order truncation to a single channel, known as \textit{Distorted Wave Born Approximation} (DWBA) and the inclusion of relevant low-lying excited states, is studied and a comparison of this results with the CC ones will be shown. 

\section{Experiment and data reduction}

\begin{figure}
	\includegraphics[width=0.5\textwidth]{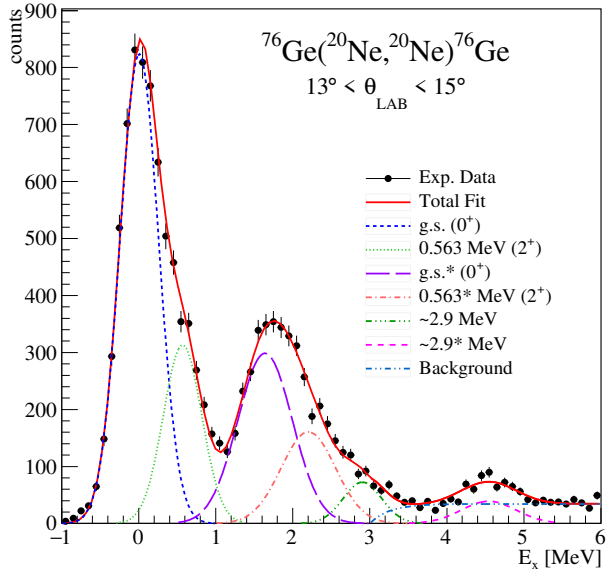}\caption{Excitation energy spectrum of \textsuperscript{76}Ge for the \textsuperscript{20}Ne~+~\textsuperscript{76}Ge elastic scattering at 306 MeV bombarding energy and \ang{13}$<\theta$\textsubscript{lab}$<$\ang{15}. Some peaks are identified in the figure by lines obtained fitting experimental data. Several states are expected to be populated starting from 3 MeV and are summarized in the fit by a unique background curve. In the legend the curves marked by an asterisk correspond to states in which the \textsuperscript{20}Ne is in the 2\textsuperscript{+} state at 1.634 MeV. }\label{spettro}
\end{figure}

A beam of \textsuperscript{20}Ne\textsuperscript{4+} ions was accelerated to 306 MeV by the INFN-LNS superconducting cyclotron \cite{rifuggiato2004}, fully stripped by crossing a thin carbon foil and transported to the target position, located at the object point of the MAGNEX large acceptance magnetic spectrometer \cite{magnex_review}. The target consisted of a 390 $\pm$ 40 $\mu$g/cm\textsuperscript{2} \textsuperscript{76}Ge layer 98\% enriched, evaporated onto a 56 $\pm$ 6 $\mu$g/cm\textsuperscript{2} Carbon backing. A copper Faraday cup of 0.8 cm entrance diameter and 3 cm depth mounted 15 cm downstream of the target was used in order to stop the beam and collect its charge. An electron suppressor polarized at -200 V and a low noise charge integrator allowed to keep the charge collection accuracy better than 10\% in all the experiment runs. 

The \textsuperscript{20}Ne ejectiles, produced in the projectile-target collisions, were momentum analysed in five different runs in which the optical axis of MAGNEX was oriented, compared to the beam direction, at $\theta$\textsubscript{opt} = \ang{8}, \ang{13}, \ang{16} and \ang{19}. In four of them, at each angle, the MAGNEX solid angle acceptance was set to 50 msr by means of slits at the entrance of the spectrometer. At $\theta$\textsubscript{opt} = \ang{8}, due to the high elastic cross section at forward angles, the beam current was optimized at about 100 epA where the signal from the Faraday cup was small compared to the electronic noise. So, in order to re-normalize the cross section with a good Faraday cup measurement, in a second run at the same  $\theta$\textsubscript{opt} the solid angle was reduced to 32 msr excluding the forward angles and increasing the beam current to measurable values. The beam current was optimized at each angular setting up to 10 enA in order to account for the strong dependence of the elastic scattering yield as a function of the ejectile angle. Under these conditions, an overlap of about $\ang{6}$ in the polar angle in the laboratory reference frame was achieved between adjacent runs. The overall measured angular range in the centre-of-mass framework was $\ang{5}$ $\le$ $\theta$\textsubscript{c.m.} $\le$ $\ang{22}$.  The magnetic fields of the MAGNEX quadrupole and bending magnet, were set in order to transport the \textsuperscript{20}Ne\textsuperscript{10+} ions corresponding to elastic scattering events at the center of the focal plane detector (FPD) \cite{cavallaro2012}. 

The data reduction strategy, including position calibration of the FPD, identification of the ejectiles of interest and reconstruction of the momentum vector at the target by inversion of the transport equations following the same method presented in previous publications  \cite{cappuzzello2010,cappuzzello2011, calabrese2018, cavallaro2019}. This procedure also allows an accurate determination of the overall detection efficiency, as presented in ref. \cite{cavallaro2011}, fundamental to extract the absolute cross section from the collected event yields. 

In the present experimental conditions the achieved angle and energy resolutions are $\delta\theta$\textsubscript{LAB}(FWHM) $\sim$ \ang{0.5} and $\delta$E(FWHM) $\sim$ 0.5 MeV, respectively. Fig. \ref{spettro} shows an example of excitation energy (E\textsubscript{x}) spectrum for the \textsuperscript{76}Ge(\textsuperscript{20}Ne,\textsuperscript{20}Ne)\textsuperscript{76}Ge reaction obtained in the angular region between \ang{13} and \ang{15} in the laboratory reference frame. The first observed peak is described as the superposition of the ground state (g.s.) and the first 2\textsuperscript{+} state of the \textsuperscript{76}Ge at E\textsubscript{x} = 0.563 MeV. The energy resolution is not enough to fully separate them, but the presence of the 2\textsuperscript{+} state is important in order to reproduce the shape of the experimental data. A structure is present at  E\textsubscript{x} $\simeq$ 1.6 MeV due to the sum of different states where the dominant is the 2\textsuperscript{+} excited state of projectile at E\textsubscript{x} = 1.634 MeV. The contribution at E\textsubscript{x} = 0.563 + 1.634 MeV corresponds to the excitation of the first 2\textsuperscript{+} states of both target and projectile. Another state is present at E\textsubscript{x} $\simeq$ 2.9 MeV also visible at E\textsubscript{x} $\simeq$ 2.9 + 1.634 MeV at the energy in which also the projectile is excited in its 2\textsuperscript{+} first excited state. The width of this state is compatible with the 0\textsuperscript{+} state of \textsuperscript{76}Ge at 2.897 MeV already observed in ($p$, $p'$) and ($\alpha$, $\alpha'$) reactions \cite{singh1995}. A multiple fit procedure was performed to describe the energy spectrum at several angles (see the example in Fig. \ref{spettro}) where the width of eachas peak was fixed according to the achieved energy resolution and the doppler enlargement due to the in flight decay of the projectile populated states. The contribution over 3 MeV in the excitation energy spectrum due to the weakly populated states of projectile and target is summarized in the fit procedure as a unique background curve. 

\begin{figure}
\includegraphics[width=0.5\textwidth]{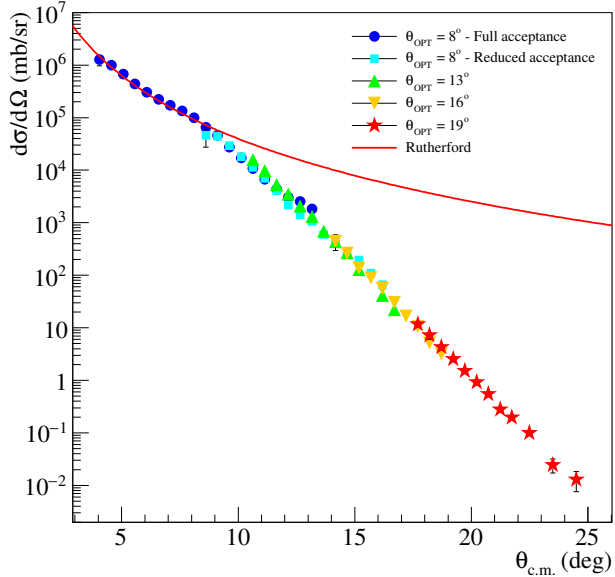}\caption{Angular distribution of differential cross section for \textsuperscript{20}Ne + \textsuperscript{76}Ge elastic scattering at 306 MeV bombarding energy. Colour points show data acquired in five separate runs for different angular settings (see text). The red line represents the Rutherford cross section.}\label{cross_sets}
\end{figure}

Experimental angular distribution data for the different angular ranges are presented in Fig. \ref{cross_sets}. The statistical error and the uncertainties coming from the fitting procedure and from the differential solid angle evaluation are included in the error bars. A scale factor equal to 1.11 was applied to all the data to ensure a good agreement with the Rutherford cross section at very forward angles. This scale factor is compatible with the estimated systematic errors in the total charge collected by the Faraday Cup and in the total number of scattering centres.     

\begin{figure}
	\includegraphics[width=0.45\textwidth]{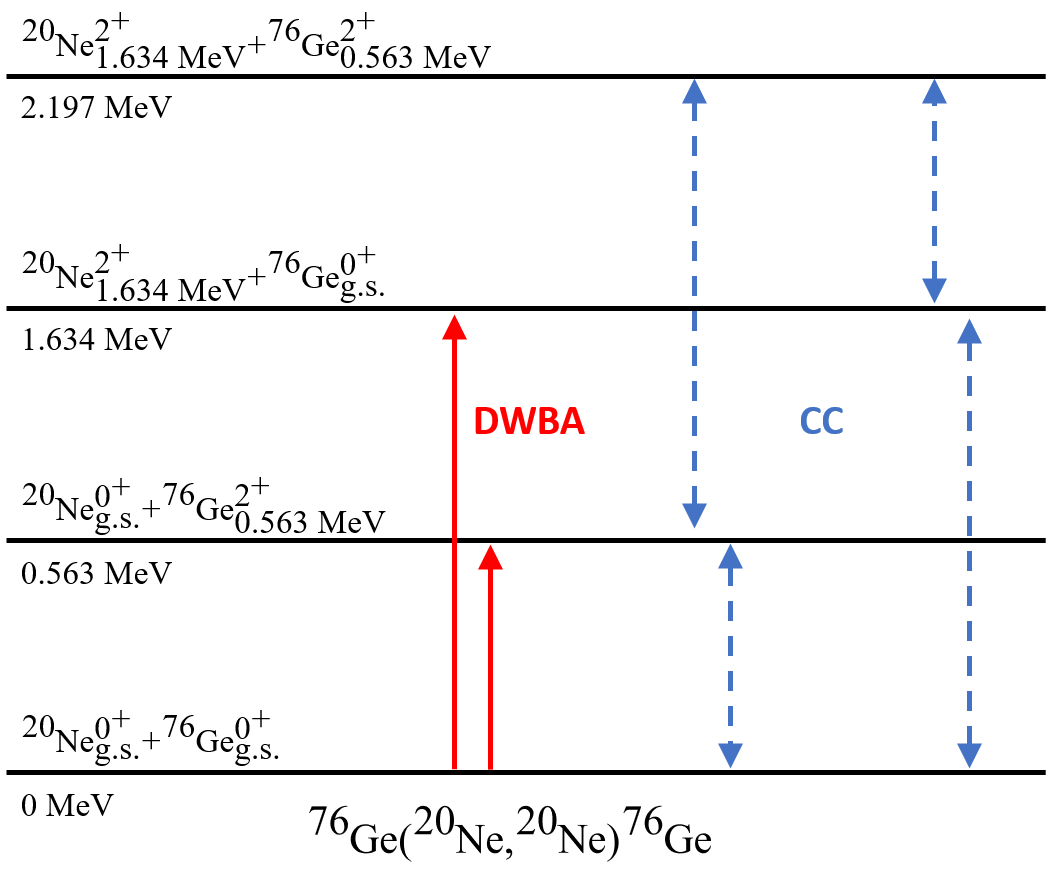}\caption{Scheme of all the coupling potentials included in the DWBA and CC calculations among the different states of projectile and target.}\label{schema_accoppiamenti}
\end{figure}

The 2\textsuperscript{+} excited states of projectile (1.634 MeV) and target (0.563 MeV) were included in the coupling scheme shown in Fig. \ref{schema_accoppiamenti}. The representation of the cross section in terms of the ratio to the Rutherford one (shown in Fig. \ref{OM_comparison} and \ref{m3y_CC}), reveals a Fresnel-like scattering pattern, as expected for such heavy colliding nuclei (Sommerfeld parameter $\eta$ = 12.9; grazing angular momentum L\textsubscript{G} $\simeq$ 129 $\hbar$). The Coulomb field dominates the scattering up to the grazing angle, located at about $\theta$\textsubscript{G} $\simeq$ \ang{9.4} in the c.m. reference frame, where the data show a Coulomb rainbow pattern. Beyond that angle the data are more sensitive to the nuclear component of the nucleus-nucleus potential, showing the typical fall-off associated to near-side and far-side scattering amplitudes. Differential cross section angular distributions have been extracted also for the first 2\textsuperscript{+} states of projectile, target and for the simultaneous excitation of both of them and are shown in Fig.~\ref{inelastic_CC}. 

\section{Theoretical analysis}

The theoretical interpretation of angular distributions for the elastic and inelastic channels of the \textsuperscript{20}Ne + \textsuperscript{76}Ge collision was performed using FRESCO code \cite{thompson1988}. The influence of the choice of different types of optical potentials was investigated comparing the calculations obtained in the optical model (OM) with the angular distribution of elastic differential cross section. The inelastic differential cross sections can be obtained from DWBA calculations. The effect and strength of the coupling are evaluated by comparing these elastic and inelastic cross sections with those obtained by the CC technique.

\subsection{Elastic Scattering (Study of the Optical Potential)}\label{elastic}

Usually, the elastic scattering can be described using a complex optical potential 
\begin{equation}
U_{OPT}(r) = V(r) +iW(r)
\end{equation} 
in which the imaginary part $W(r)$ summarizes the non-elastic contributions in the average nucleus-nucleus interaction and results in an absorptive component for the elastic cross section. In the present work three different optical potentials are tested: the parametric Akyuz-Whinter (AW) \cite{akyuzwinther, broglia1991} and two double folding optical potentials, the DFOL  \cite{lenske_dfol} and  the S\~ao Paulo Potential (SPP) \cite{chamon2002, pereira2009}.

The real part $V(r)$ of the AW potential is a central Woods-Saxon function in which radius $R$, strength $V_{0}$ and diffuseness $a$ are extracted by the interpolation of double folding potentials fitted on many nuclear systems in a quite large range of masses and energies. In the double folding approach, $V$ is obtained by the folding of the nucleon-nucleon interaction $V_{NN}(\mathbf{r}_{1},\mathbf{r}_{2}, E)$ on the densities of the two involved nuclei $\rho_1(\mathbf{r}_1)$ and $\rho_2(\mathbf{r}_2)$:
\begin{equation}\label{fold_potential}
V(r)=\int d\mathbf{r}_1 d\mathbf{r}_2\rho_1(\mathbf{r}_1)\rho_2(\mathbf{r}_2)V_{NN}(\mathbf{r}_1,\mathbf{r}_2, E)
\end{equation} 
where $r$ is the projectile-target distance and $E$ is the energy per nucleon in the centre-of-mass reference frame. 

The densities $\rho_{j}(\mathbf{r}_{j})$ of projectile and target used in the foldings of both SPP and DFOL potentials are parametrized by Wood-Saxon profiles. Density parameters used in the SPP and DFOL folding are taken from ref. \cite{chamon2002} and ref. \cite{lenske_dfol} respectively. In the DFOL approach the T-matrix interaction at E\textsubscript{lab} = 50 MeV from ref. \cite{franey1985} was used to calculate the real and the imaginary part of the optical potential through the same procedure described in ref. \cite{cappuzzello2004}.  The imaginary parts of the AW and SPP optical potentials were obtained from the real ones scaling their strength by a factor 0.78 as, also seen in previous works of ref. \cite{ermamatov2016,carbone2017, ermamatov2017, paes2017},  and a 0.8 scale factor was applied also to the DFOL optical potential.

\begin{figure}
	\includegraphics[width=0.5\textwidth]{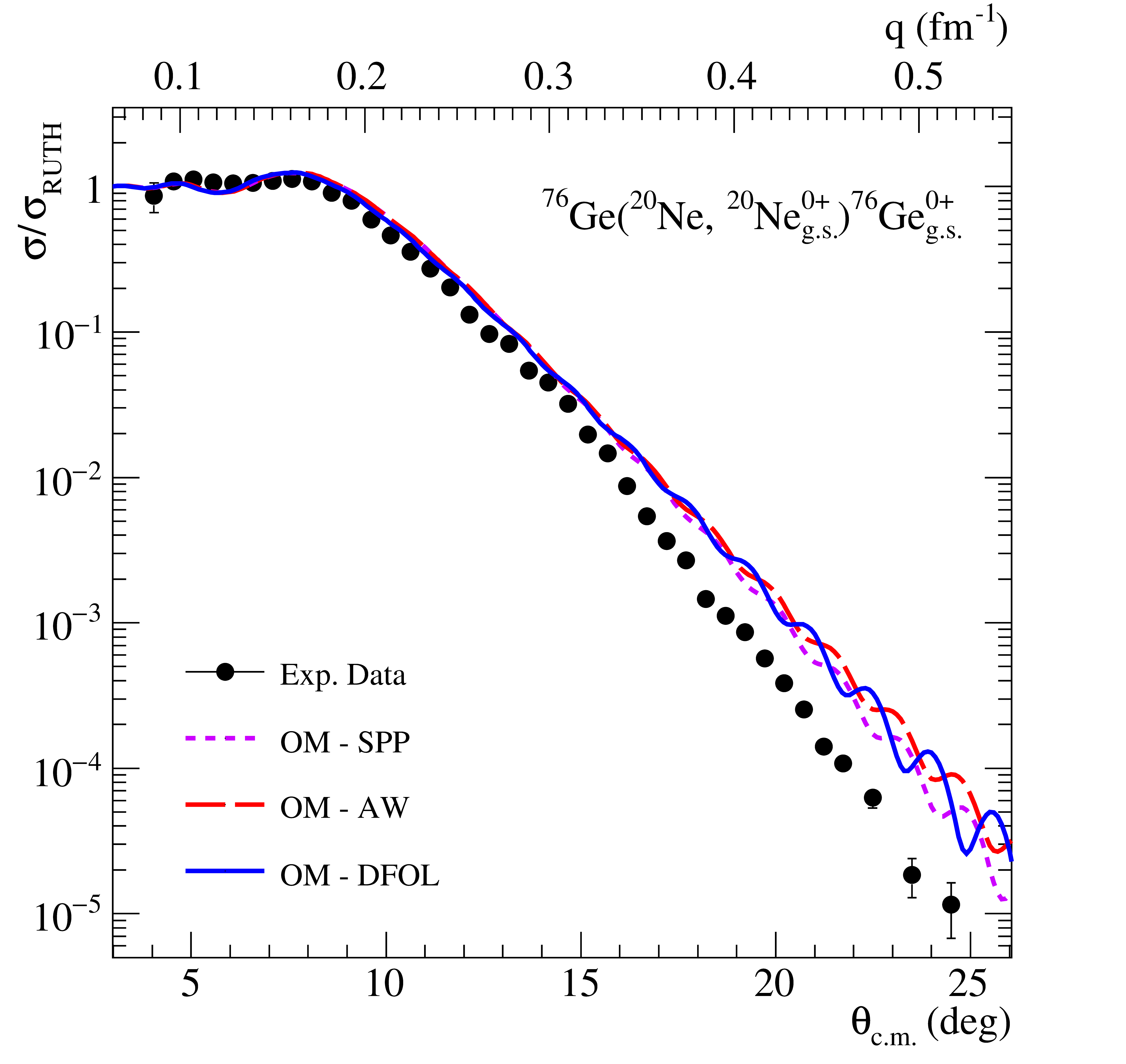}\caption{Angular distribution of the elastic differential cross section in terms of its ratio with the Rutherford cross section $\sigma_{RUTH}$. Black dots are obtained from the coloured ones in Fig. \ref{cross_sets} by an average weighted on the error of each measurement. Lines represent optical model calculations performed on the context of different optical potentials. }\label{OM_comparison}
\end{figure}

The results of OM calculations performed with the three tested optical potentials and the experimental data in the $\sigma$/$\sigma$\textsubscript{RUTH} representation are shown in Fig. \ref{OM_comparison}. There is not a significant difference between the three calculations. This fact confirms that the theoretical description of elastic scattering is not strongly dependent on the choice of the optical potential since the strong absorption confines the reaction source on the surface of the colliding systems. Although different in terms of root main square radius and volume integral per nucleon, the total reaction cross section (see Tab.\ref{tabella_prop_pot}) and the angular distribution (see Fig.\ref{OM_comparison}) is very similar for all the studied potentials.

In Fig. \ref{OM_comparison} the experimental data beyond the grazing angle ($\sim$ \ang{9.4}) show a slope steeper than that obtained from our OM calculations. Since all of them, performed with different potentials, are in agreement with each other, the discrepancy could indicate a common drawback in the description of the geometrical properties of the nuclear densities. An important approximation is that the density profiles used in the folding of the DFOL and SPP optical potentials for both projectile and target isotopes that are assumed to be spherical. Moreover, the AW optical potential parameters are obtained from interpolations of double folding optical potentials of several nuclear systems in a quite large range of energies and masses. Also in this case the nuclear systems are assumed to be spherical. 

Since both g.s. quadrupole moments of projectile and target are large \cite{prytichenko_data_table}, one has to infer that \textsuperscript{20}Ne and \textsuperscript{76}Ge isotopes are significantly deformed. Moreover at these transferred momenta the hypothesis of \textit{frozen} nuclear spherical matter in the reaction mechanism is at limit. An effective way to take these arguments into account in the building of the optical potential is a changing of the density matter profiles used in the folding of the NN interactions, as also described in refs. \cite{fonseca2019, crema2011, crema2018}. Nevertheless, a re-normalization is mandatory in order to keep constant the volume integral of the nuclear densities to fix the number of nucleons.

\begin{figure*}\centering
	\subfigure[][DFOL Potential]{\includegraphics[width=0.48\textwidth]{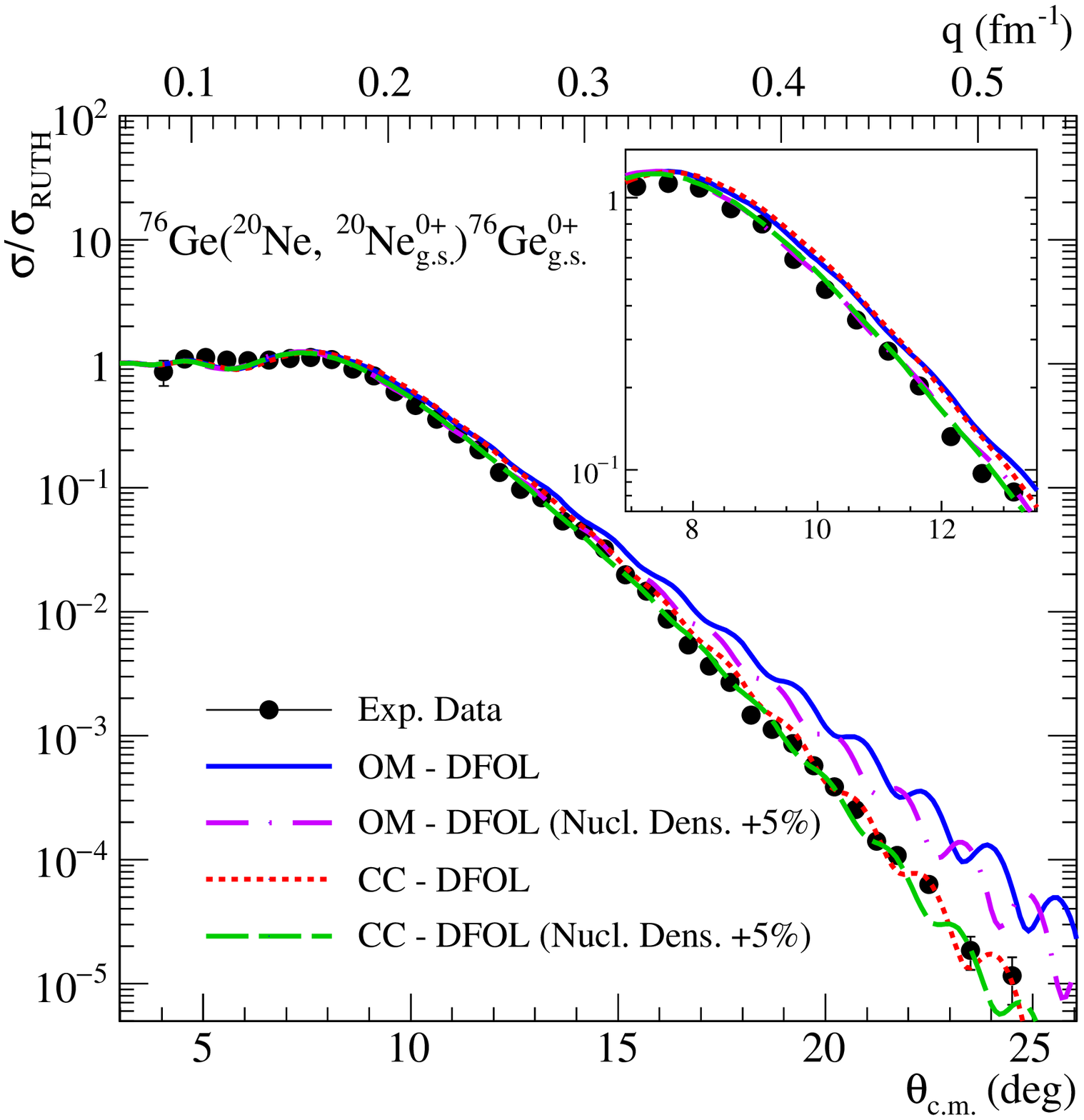}}
	\subfigure[][SPP Potential]{\includegraphics[width=0.48\textwidth]{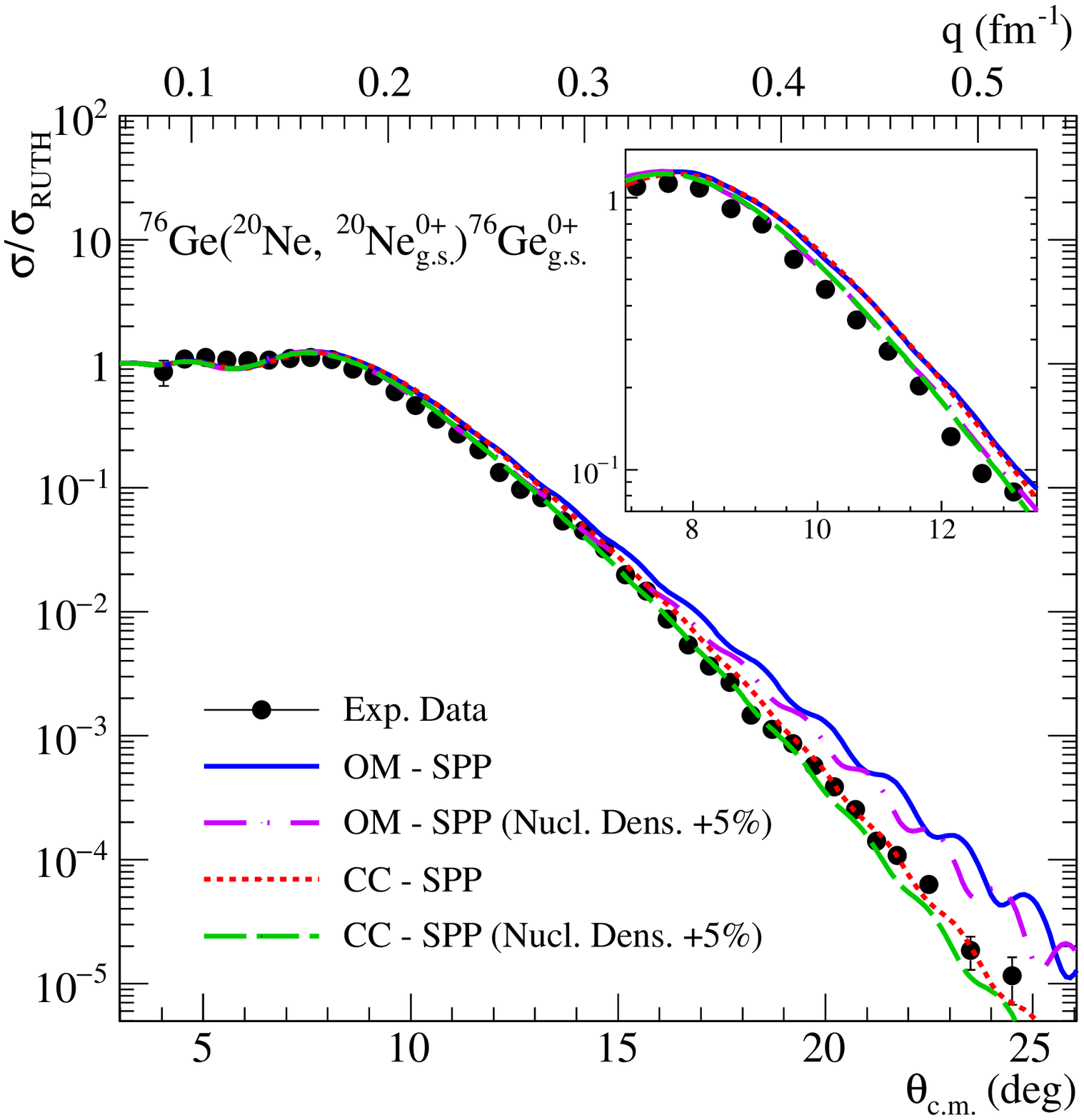}}
	\caption{Elastic scattering angular distribution of differential cross section in terms of the $\sigma$/$\sigma_{ruth}$ ratio. The (a) and (b) panels show the results obtained in OM and CC approaches with the DFOL and SPP double folding potentials respectively. In both panels, the solid blue and the dot-dashed violet curves represent the OM calculations with the standard potentials and by 5\% the radius of the nuclear density distributions respectively (see text). The red dotted and the green dashed curves are the results of the CC approach obtained with the standard and the modified potentials.}\label{m3y_CC}
\end{figure*}

In the present work this operation has been performed increasing by 5\% the radius of the nuclear density profiles in the folding of both the DFOL and the SPP optical potentials and renormalizing the central density parameter. The comparison between the standard approach and the new one is shown in Fig. \ref{m3y_CC} by the solid blue and dot-dashed violet curves. The increase of radius appears important to correctly describe the experimental shape up to about \ang{14}, where geometrical properties of nuclei are more relevant. However the change of slope observed in the experimental data above \ang{14} is still not described in the OM framework.

\begin{table}\renewcommand\arraystretch{1.4}\caption{Total reaction cross sections $\sigma_{R}$, volume integral per nucleon $J$ and root mean square radii $\sqrt{\braket{R^2}}$ for the real ($V$) and imaginary ($W$) part of the AW, DFOL and SPP potentials.}\label{tabella_prop_pot}
	\begin{tabular}{p{40pt}rcccc}
		\hline
		\multirow{2}{*}{}	&$\sigma_R$		&$J_V$	&$J_W$	&$\sqrt{\braket{R^2_V}}$	&$\sqrt{\braket{R^2_W}}$\\
		 					&(\textit{mb})	&(\textit{MeV fm\textsuperscript{3}})	& (\textit{MeV fm\textsuperscript{3}})	&(\textit{fm})	& (\textit{fm})\\
		 
		\hline
		AW	&2807 &-1390	&-1042	&7.03	&7.03\\
		DFOL&2819 &-6928	&-5108	&5.30	&5.18\\
		SPP &2799 &-5426	&-4069	&5.32	&5.32\\
		\hline
	\end{tabular}
\end{table}

The effect in the elastic channel of the inclusion of couplings with the first low-lying excited states of projectile and target is shown in Fig. \ref{m3y_CC} by the red-dotted and green dashed curves for both the double folding potentials (DFOL and SPP). This effect is here evaluated also in terms of the transferred momentum $q$, calculated by the approximated expression $\sim 2E_{c.m.}/\hbar c\sin\theta_{c.m.}$. The effect of couplings starts to be important in the description of the elastic scattering over \ang{15}, at almost 0.3 fm\textsuperscript{-1}, where the absolute cross sections for the elastic and the inelastic scattering channels become comparable. Other details on the performed CC calculations are described in the next subsection.

\subsection{Inelastic Scattering (Study of coupling contributions)}

\begin{figure}
	\includegraphics[width=0.5\textwidth]{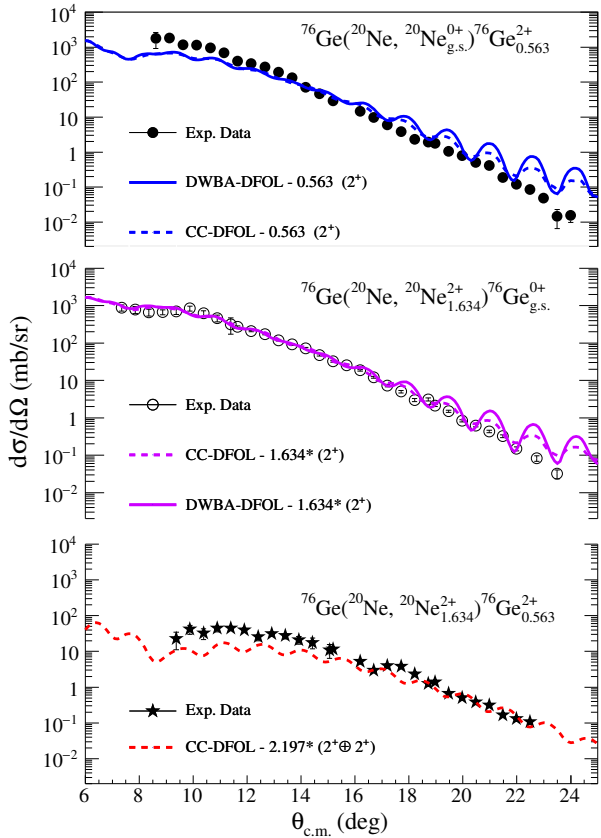}\caption{Angular distribution of differential cross section for the low-lying states of projectile and target populated in the \textsuperscript{20}Ne + \textsuperscript{76}Ge inelastic scattering at 306 MeV bombarding energy: in the top panel 2\textsuperscript{+} state of \textsuperscript{76}Ge at 0.563 MeV; in the middle panel the 2\textsuperscript{+} state of the \textsuperscript{20}Ne at 1.634 MeV; in the bottom panel 2.197 MeV state that corresponds to the excitation of both projectile and target 2\textsuperscript{+} $\oplus$ 2\textsuperscript{+}. Solid lines are the results obtained in DWBA approach; dashed lines are CC calculations.}\label{inelastic_CC}
\end{figure}

The angular distributions of the transition to the first excited states of the projectile (2\textsuperscript{+} at 1.634 MeV), target (2\textsuperscript{+} at 0.563 MeV) and for the simultaneous excitation of both (2\textsuperscript{+}$\oplus$2\textsuperscript{+} at 2.197 MeV) are shown in Fig. \ref{inelastic_CC}. Calculations for the excited states were performed in DWBA and CC approaches in the context of a rotational model in which the 2\textsuperscript{+} states are treated as quadrupole excitations of pure rotors. Coulomb deformations of nuclei were introduced in terms of reduced transition probabilities. B(E2; 0\textsuperscript{+}$\rightarrow$2\textsuperscript{+}) = 0.0333 e\textsuperscript{2}b\textsuperscript{2} for \textsuperscript{20}Ne and B(E2; 0\textsuperscript{+}$\rightarrow$2\textsuperscript{+}) = 0.2735 e\textsuperscript{2}b\textsuperscript{2} for \textsuperscript{76}Ge are obtained from ref. \cite{prytichenko_data_table} and were used to describe Coulomb deformations of both projectile and target. The $V_2^i(r)$ nuclear coupling potentials are treated in the first-order approximation described in ref. \cite{satchler} through the following formula
\begin{equation}
V_{2}^{i}(r) = - \frac{\delta_{2}^{i}}{\sqrt{4\pi}} \frac{dU(r)}{dr}
\end{equation}
where the deformation lengths $\delta_2^i$=$\beta_2^i R^i$ are calculated through the deformations $\beta_2^i$ and  radii $R_i$ of each nucleus indicated by $i$. For the imaginary coupling potentials, the same radial form factors are assumed taking the $\beta_2^{real}$ = $\beta_2^{imag}$ convention. Calculations shown in Fig. \ref{inelastic_CC} have been performed using only the DFOL optical potential since the results obtained with the SPP are in complete agreement with them. The DFOL Optical Potential is the one obtained increasing the radius of Woods-Saxon nuclear density profiles of 5\% to ensure a good agreement with elastic scattering data at forward angles, as previously described in section \ref{elastic}. 

Calculations for the first low-lying excited states are in a reasonable agreement with experimental data. The effect of higher order terms of couplings to the elastic channel is not relevant in the description of the average shape of the angular distributions for the inelastic transitions, as it is shown in the top and in the middle panels of Fig. \ref{inelastic_CC}. Here the main effect of couplings is a smoothing in the oscillation pattern over \ang{16} also present in the data. A description of the angular distribution for the 2\textsuperscript{+}$\oplus$2\textsuperscript{+} state at 2.197 MeV is not possible in the DWBA one-step approach since the relative transition is a second order process (see Fig. \ref{schema_accoppiamenti}). The CC calculation performed is shown in the bottom panel of Fig. \ref{inelastic_CC}, and the obtained result is in good agreement with the experimental data. 

\section{Conclusions}

In the present work, elastic and inelastic scattering of the \textsuperscript{20}Ne + \textsuperscript{76}Ge at 306 MeV were studied for the first time, including good theoretical descriptions of their excitations by using Optical Model and CC  approaches. The g.s. to g.s. transition was separated by the other inelastic channels thanks to the very good resolution gained through a careful tuning of the experimental set-up and the applied advanced analysis. Moreover, the small error bars and the overall quality of the experimental data are such as to justify the attempt of using sophisticated microscopic analysis.

The capability of several optical potentials to fit the angular distribution of the measured differential cross sections was tested showing that the response with the AW optical potential and the SPP and DFOL double folding potentials is practically the same. Standard versions of all these potentials were not enough to correctly describe the experimental data above the grazing angles. This goal was partially achieved by working on the geometrical parameters of the nuclear matter densities of isotopes involved. Since both the \textsuperscript{20}Ne and \textsuperscript{76}Ge are deformed, an effective way to take into account this property is to increase the radius of their nuclear matter densities of 5\% obtaining a good agreement until about \ang{14} in the elastic scattering angular distribution. 

At transferred momentum higher than 0.3 fm\textsuperscript{-1} the effect of couplings with first low-lying excited states becomes essential to describe especially the elastic scattering where the cross section becomes comparable with the ones of first 2\textsuperscript{+} states. Other studies have to be performed in order to check if contributions from coupling to the projectile and target 2\textsuperscript{+} low-lying states are relevant to the DCE and SCE reaction cross sections. In this context, new analysis have to be performed in order to find the appropriate characterization of the model space and average interaction for the outgoing partition. 

\section*{Acknowledgment}
This project has received funding from the European Research Council (ERC) under the European Union's Horizon 2020 research and innovation programme (grant agreement No. 714625) and from the Mexican grants DGAPA-PAPIIT IG101016, IA103218, PIIF2018 and CONACyT 294537. J.A.L. acknowledges funding from the European Union's Horizon 2020 research and innovation
programme under grant agreement N. 654002, and from the Spanish Ministerio de Economia y Competitividad and FEDER funds under Project FIS2017-88410-P.    

\bibliography{biblio}

\begin{thebibliography}{46}
\expandafter\ifx\csname natexlab\endcsname\relax\def\natexlab#1{#1}\fi
\expandafter\ifx\csname bibnamefont\endcsname\relax
  \def\bibnamefont#1{#1}\fi
\expandafter\ifx\csname bibfnamefont\endcsname\relax
  \def\bibfnamefont#1{#1}\fi
\expandafter\ifx\csname citenamefont\endcsname\relax
  \def\citenamefont#1{#1}\fi
\expandafter\ifx\csname url\endcsname\relax
  \def\url#1{\texttt{#1}}\fi
\expandafter\ifx\csname urlprefix\endcsname\relax\def\urlprefix{URL }\fi
\providecommand{\bibinfo}[2]{#2}
\providecommand{\eprint}[2][]{\url{#2}}

\bibitem[{\citenamefont{Dell'Oro et~al.}(2014)\citenamefont{Dell'Oro, Marcocci,
  and Vissani}}]{delloro2014}
\bibinfo{author}{\bibfnamefont{S.}~\bibnamefont{Dell'Oro}},
  \bibinfo{author}{\bibfnamefont{S.}~\bibnamefont{Marcocci}}, \bibnamefont{and}
  \bibinfo{author}{\bibfnamefont{F.}~\bibnamefont{Vissani}},
  \bibinfo{journal}{Phys. Rev. D} \textbf{\bibinfo{volume}{90}},
  \bibinfo{pages}{033005} (\bibinfo{year}{2014}),
  \urlprefix\url{https://link.aps.org/doi/10.1103/PhysRevD.90.033005}.

\bibitem[{\citenamefont{Cappuzzello et~al.}(2018)}]{numen2018}
\bibinfo{author}{\bibfnamefont{F.}~\bibnamefont{Cappuzzello}}
  \bibnamefont{et~al.}, \bibinfo{journal}{Eur. Phys. J. A}
  \textbf{\bibinfo{volume}{54}}, \bibinfo{pages}{72} (\bibinfo{year}{2018}),
  \urlprefix\url{https://doi.org/10.1140/epja/i2018-12509-3}.

\bibitem[{\citenamefont{Cappuzzello et~al.}(2015)}]{numen_proc_braz}
\bibinfo{author}{\bibfnamefont{F.}~\bibnamefont{Cappuzzello}}
  \bibnamefont{et~al.}, \bibinfo{journal}{J. Phys.: Conf. Ser}
  \textbf{\bibinfo{volume}{630}}, \bibinfo{pages}{012018}
  (\bibinfo{year}{2015}),
  \urlprefix\url{https://doi.org/10.1088%2F1742-6596%2F630%2F1%2F012018}.

\bibitem[{\citenamefont{Matsubara et~al.}(2013)}]{matsubara_riken}
\bibinfo{author}{\bibfnamefont{H.}~\bibnamefont{Matsubara}}
  \bibnamefont{et~al.}, \bibinfo{journal}{Few-Body Systems}
  \textbf{\bibinfo{volume}{54}}, \bibinfo{pages}{1433} (\bibinfo{year}{2013}),
  ISSN \bibinfo{issn}{0177-7963}.

\bibitem[{\citenamefont{Kisamori et~al.}(2016)}]{kisamori_riken}
\bibinfo{author}{\bibfnamefont{K.}~\bibnamefont{Kisamori}}
  \bibnamefont{et~al.}, \bibinfo{journal}{Phys. Rev. Lett.}
  \textbf{\bibinfo{volume}{116}}, \bibinfo{pages}{052501}
  (\bibinfo{year}{2016}),
  \urlprefix\url{https://link.aps.org/doi/10.1103/PhysRevLett.116.052501}.

\bibitem[{\citenamefont{{Takahisa} et~al.}(2017)\citenamefont{{Takahisa},
  {Ejiri}, {Akimune}, {Fujita}, {Matumiya}, {Ohta}, {Shima}, {Tanaka}, and
  {Yosoi}}}]{takahisa_RNCP}
\bibinfo{author}{\bibfnamefont{K.}~\bibnamefont{{Takahisa}}},
  \bibinfo{author}{\bibfnamefont{H.}~\bibnamefont{{Ejiri}}},
  \bibinfo{author}{\bibfnamefont{H.}~\bibnamefont{{Akimune}}},
  \bibinfo{author}{\bibfnamefont{H.}~\bibnamefont{{Fujita}}},
  \bibinfo{author}{\bibfnamefont{R.}~\bibnamefont{{Matumiya}}},
  \bibinfo{author}{\bibfnamefont{T.}~\bibnamefont{{Ohta}}},
  \bibinfo{author}{\bibfnamefont{T.}~\bibnamefont{{Shima}}},
  \bibinfo{author}{\bibfnamefont{M.}~\bibnamefont{{Tanaka}}}, \bibnamefont{and}
  \bibinfo{author}{\bibfnamefont{M.}~\bibnamefont{{Yosoi}}},
  \bibinfo{journal}{arXiv e-prints} \bibinfo{eid}{arXiv:1703.08264}
  (\bibinfo{year}{2017}), \eprint{1703.08264}.

\bibitem[{\citenamefont{Shimizu et~al.}(2018)\citenamefont{Shimizu,
  Men{\'e}ndez, and Yako}}]{shimizu}
\bibinfo{author}{\bibfnamefont{N.}~\bibnamefont{Shimizu}},
  \bibinfo{author}{\bibfnamefont{J.}~\bibnamefont{Men{\'e}ndez}},
  \bibnamefont{and} \bibinfo{author}{\bibfnamefont{K.}~\bibnamefont{Yako}},
  \bibinfo{journal}{Phys. Rev. Lett.} \textbf{\bibinfo{volume}{120}},
  \bibinfo{pages}{142502} (\bibinfo{year}{2018}), \eprint{1709.01088}.

\bibitem[{\citenamefont{Lenske et~al.}(2018)\citenamefont{Lenske, Bellone,
  Colonna, and Lay}}]{lenske2018}
\bibinfo{author}{\bibfnamefont{H.}~\bibnamefont{Lenske}},
  \bibinfo{author}{\bibfnamefont{J.~I.} \bibnamefont{Bellone}},
  \bibinfo{author}{\bibfnamefont{M.}~\bibnamefont{Colonna}}, \bibnamefont{and}
  \bibinfo{author}{\bibfnamefont{J.~A.} \bibnamefont{Lay}},
  \bibinfo{journal}{Phys. Rev.} \textbf{\bibinfo{volume}{C98}},
  \bibinfo{pages}{044620} (\bibinfo{year}{2018}), \eprint{1803.06290}.

\bibitem[{\citenamefont{Lenske}(2018)}]{lenske_CNNP}
\bibinfo{author}{\bibfnamefont{H.}~\bibnamefont{Lenske}}, \bibinfo{journal}{J.
  Phys.: Conf. Ser.} \textbf{\bibinfo{volume}{1056}}, \bibinfo{pages}{012030}
  (\bibinfo{year}{2018}),
  \urlprefix\url{https://doi.org/10.1088%2F1742-6596%2F1056%2F1%2F012030}.

\bibitem[{col(2019)}]{colonna}
\emph{\bibinfo{title}{{Heavy ion charge exchange reactions and the link with
  beta decay processes}}}, vol.~\bibinfo{volume}{1} (\bibinfo{publisher}{CERN},
  \bibinfo{address}{Geneva}, \bibinfo{year}{2019}), ISBN
  \bibinfo{isbn}{9789290835196, 9789290835202}.

\bibitem[{\citenamefont{Kumar et~al.}(2017)\citenamefont{Kumar, Kanungo, Calci,
  Navratil, Sanetullaev, Alcorta, Bildstein, Christian, Davids
  et~al.}}]{kumar2017}
\bibinfo{author}{\bibfnamefont{A.}~\bibnamefont{Kumar}},
  \bibinfo{author}{\bibfnamefont{R.}~\bibnamefont{Kanungo}},
  \bibinfo{author}{\bibfnamefont{A.}~\bibnamefont{Calci}},
  \bibinfo{author}{\bibfnamefont{P.}~\bibnamefont{Navratil}},
  \bibinfo{author}{\bibfnamefont{A.}~\bibnamefont{Sanetullaev}},
  \bibinfo{author}{\bibfnamefont{M.}~\bibnamefont{Alcorta}},
  \bibinfo{author}{\bibfnamefont{V.}~\bibnamefont{Bildstein}},
  \bibinfo{author}{\bibfnamefont{G.}~\bibnamefont{Christian}},
  \bibinfo{author}{\bibfnamefont{B.}~\bibnamefont{Davids}},
  \bibnamefont{et~al.}, \bibinfo{journal}{Phys. Rev. Lett.}
  \textbf{\bibinfo{volume}{118}}, \bibinfo{pages}{262502}
  (\bibinfo{year}{2017}),
  \urlprefix\url{https://link.aps.org/doi/10.1103/PhysRevLett.118.262502}.

\bibitem[{\citenamefont{Satchler}(1983)}]{satchler}
\bibinfo{author}{\bibfnamefont{G.}~\bibnamefont{Satchler}},
  \emph{\bibinfo{title}{Direct nuclear reactions}}, vol.~\bibinfo{volume}{68}
  of \emph{\bibinfo{series}{International series of monographs on physics}}
  (\bibinfo{publisher}{Clarendon Press}, \bibinfo{address}{Oxford, UK},
  \bibinfo{year}{1983}), ISBN \bibinfo{isbn}{9780198512691},
  \urlprefix\url{https://books.google.it/books?id=\_SQFAQAAIAAJ}.

\bibitem[{\citenamefont{Satchler and Love}(1979)}]{satchler1979}
\bibinfo{author}{\bibfnamefont{G.}~\bibnamefont{Satchler}} \bibnamefont{and}
  \bibinfo{author}{\bibfnamefont{W.}~\bibnamefont{Love}},
  \bibinfo{journal}{Physics Reports} \textbf{\bibinfo{volume}{55}},
  \bibinfo{pages}{183 } (\bibinfo{year}{1979}), ISSN \bibinfo{issn}{0370-1573},
  \urlprefix\url{http://www.sciencedirect.com/science/article/pii/0370157379900814}.

\bibitem[{\citenamefont{Sgouros et~al.}(2013)}]{sgouros2013}
\bibinfo{author}{\bibfnamefont{O.}~\bibnamefont{Sgouros}} \bibnamefont{et~al.},
  \bibinfo{journal}{International Journal of Modern Physics E}
  \textbf{\bibinfo{volume}{22}}, \bibinfo{pages}{50073} (\bibinfo{year}{2013}).

\bibitem[{\citenamefont{Trzci\ifmmode~\acute{n}\else \'{n}\fi{}ska
  et~al.}(2015)}]{trz2015}
\bibinfo{author}{\bibfnamefont{A.}~\bibnamefont{Trzci\ifmmode~\acute{n}\else
  \'{n}\fi{}ska}} \bibnamefont{et~al.}, \bibinfo{journal}{Phys. Rev. C}
  \textbf{\bibinfo{volume}{92}}, \bibinfo{pages}{034619}
  (\bibinfo{year}{2015}),
  \urlprefix\url{https://link.aps.org/doi/10.1103/PhysRevC.92.034619}.

\bibitem[{\citenamefont{Ohkubo and Hirabayashi}(2014)}]{ohkubo2014}
\bibinfo{author}{\bibfnamefont{S.}~\bibnamefont{Ohkubo}} \bibnamefont{and}
  \bibinfo{author}{\bibfnamefont{Y.}~\bibnamefont{Hirabayashi}},
  \bibinfo{journal}{Phys. Rev. C} \textbf{\bibinfo{volume}{89}},
  \bibinfo{pages}{051601(R)} (\bibinfo{year}{2014}),
  \urlprefix\url{https://link.aps.org/doi/10.1103/PhysRevC.89.051601}.

\bibitem[{\citenamefont{Ohkubo and Yamashita}(2002)}]{ohkubo2002}
\bibinfo{author}{\bibfnamefont{S.}~\bibnamefont{Ohkubo}} \bibnamefont{and}
  \bibinfo{author}{\bibfnamefont{K.}~\bibnamefont{Yamashita}},
  \bibinfo{journal}{Phys. Rev. C} \textbf{\bibinfo{volume}{66}},
  \bibinfo{pages}{021301(R)} (\bibinfo{year}{2002}),
  \urlprefix\url{https://link.aps.org/doi/10.1103/PhysRevC.66.021301}.

\bibitem[{\citenamefont{Zagatto et~al.}(2018)\citenamefont{Zagatto,
  Cappuzzello, Lubian, Cavallaro, Linares, Carbone, Agodi, Foti, Tudisco
  et~al.}}]{zagatto2018}
\bibinfo{author}{\bibfnamefont{V.}~\bibnamefont{Zagatto}},
  \bibinfo{author}{\bibfnamefont{F.}~\bibnamefont{Cappuzzello}},
  \bibinfo{author}{\bibfnamefont{J.}~\bibnamefont{Lubian}},
  \bibinfo{author}{\bibfnamefont{M.}~\bibnamefont{Cavallaro}},
  \bibinfo{author}{\bibfnamefont{R.}~\bibnamefont{Linares}},
  \bibinfo{author}{\bibfnamefont{D.}~\bibnamefont{Carbone}},
  \bibinfo{author}{\bibfnamefont{C.}~\bibnamefont{Agodi}},
  \bibinfo{author}{\bibfnamefont{A.}~\bibnamefont{Foti}},
  \bibinfo{author}{\bibfnamefont{S.}~\bibnamefont{Tudisco}},
  \bibnamefont{et~al.}, \bibinfo{journal}{Phys. Rev. C}
  \textbf{\bibinfo{volume}{97}}, \bibinfo{pages}{054608}
  (\bibinfo{year}{2018}).

\bibitem[{\citenamefont{Pereira et~al.}(2012)}]{pereira_rainbow}
\bibinfo{author}{\bibfnamefont{D.}~\bibnamefont{Pereira}} \bibnamefont{et~al.},
  \bibinfo{journal}{Phys. Lett. B} \textbf{\bibinfo{volume}{710}},
  \bibinfo{pages}{426 } (\bibinfo{year}{2012}), ISSN \bibinfo{issn}{0370-2693},
  \urlprefix\url{http://www.sciencedirect.com/science/article/pii/S0370269312003061}.

\bibitem[{\citenamefont{Oliveira et~al.}(2013)}]{oliveira_rainbow}
\bibinfo{author}{\bibfnamefont{J.~R.~B.} \bibnamefont{Oliveira}}
  \bibnamefont{et~al.}, \bibinfo{journal}{J. Phys. G: Nucl. Part. Phys.}
  \textbf{\bibinfo{volume}{40}}, \bibinfo{pages}{105101}
  (\bibinfo{year}{2013}),
  \urlprefix\url{https://doi.org/10.1088%2F0954-3899%2F40%2F10%2F105101}.

\bibitem[{\citenamefont{Cappuzzello
  et~al.}(2016{\natexlab{a}})}]{cappuzzello2016}
\bibinfo{author}{\bibfnamefont{F.}~\bibnamefont{Cappuzzello}}
  \bibnamefont{et~al.}, \bibinfo{journal}{Eur. Phys. J. A}
  \textbf{\bibinfo{volume}{52}}, \bibinfo{pages}{169}
  (\bibinfo{year}{2016}{\natexlab{a}}).

\bibitem[{\citenamefont{Akyuz and Winther}(1979)}]{akyuzwinther}
\bibinfo{author}{\bibfnamefont{R.}~\bibnamefont{Akyuz}} \bibnamefont{and}
  \bibinfo{author}{\bibfnamefont{A.}~\bibnamefont{Winther}}, in
  \emph{\bibinfo{booktitle}{Proc. Enrico Fermi Int. School of Physics}}, edited
  by \bibinfo{editor}{\bibfnamefont{R.~A.} \bibnamefont{Broglia}},
  \bibinfo{editor}{\bibfnamefont{C.~H.} \bibnamefont{Dasso}}, \bibnamefont{and}
  \bibinfo{editor}{\bibfnamefont{R.}~\bibnamefont{Ricci}}
  (\bibinfo{publisher}{North-Holland}, \bibinfo{address}{Amsterdam},
  \bibinfo{year}{1979}).

\bibitem[{\citenamefont{Broglia and Winther}(1991)}]{broglia1991}
\bibinfo{author}{\bibfnamefont{R.}~\bibnamefont{Broglia}} \bibnamefont{and}
  \bibinfo{author}{\bibfnamefont{A.}~\bibnamefont{Winther}},
  \emph{\bibinfo{title}{Heavy Ion Reactions: The elementary processes}},
  Frontiers in Physics (\bibinfo{publisher}{Addison-Wesley},
  \bibinfo{address}{University of Michigan}, \bibinfo{year}{1991}), ISBN
  \bibinfo{isbn}{9780201513929},
  \urlprefix\url{https://books.google.it/books?id=GztRAAAAMAAJ}.

\bibitem[{\citenamefont{Chamon et~al.}(2002)\citenamefont{Chamon, Carlson,
  Gasques, Pereira, De~Conti, Alvarez, Hussein, Candido~Ribeiro, Rossi, and
  Silva}}]{chamon2002}
\bibinfo{author}{\bibfnamefont{L.~C.} \bibnamefont{Chamon}},
  \bibinfo{author}{\bibfnamefont{B.~V.} \bibnamefont{Carlson}},
  \bibinfo{author}{\bibfnamefont{L.~R.} \bibnamefont{Gasques}},
  \bibinfo{author}{\bibfnamefont{D.}~\bibnamefont{Pereira}},
  \bibinfo{author}{\bibfnamefont{C.}~\bibnamefont{De~Conti}},
  \bibinfo{author}{\bibfnamefont{M.~A.~G.} \bibnamefont{Alvarez}},
  \bibinfo{author}{\bibfnamefont{M.~S.} \bibnamefont{Hussein}},
  \bibinfo{author}{\bibfnamefont{M.~A.} \bibnamefont{Candido~Ribeiro}},
  \bibinfo{author}{\bibfnamefont{E.~S.} \bibnamefont{Rossi},
  \bibfnamefont{Jr.}}, \bibnamefont{and} \bibinfo{author}{\bibfnamefont{C.~P.}
  \bibnamefont{Silva}}, \bibinfo{journal}{Phys. Rev.}
  \textbf{\bibinfo{volume}{C66}}, \bibinfo{pages}{014610}
  (\bibinfo{year}{2002}), \eprint{nucl-th/0202015}.

\bibitem[{\citenamefont{Pereira et~al.}(2009)\citenamefont{Pereira, Lubian,
  Oliveira, de~Sousa, and Chamon}}]{pereira2009}
\bibinfo{author}{\bibfnamefont{D.}~\bibnamefont{Pereira}},
  \bibinfo{author}{\bibfnamefont{J.}~\bibnamefont{Lubian}},
  \bibinfo{author}{\bibfnamefont{J.~R.~B.} \bibnamefont{Oliveira}},
  \bibinfo{author}{\bibfnamefont{D.~P.} \bibnamefont{de~Sousa}},
  \bibnamefont{and} \bibinfo{author}{\bibfnamefont{L.~C.}
  \bibnamefont{Chamon}}, \bibinfo{journal}{Phys. Lett. B}
  \textbf{\bibinfo{volume}{670}}, \bibinfo{pages}{330 } (\bibinfo{year}{2009}),
  ISSN \bibinfo{issn}{0370-2693},
  \urlprefix\url{http://www.sciencedirect.com/science/article/pii/S0370269308013543}.

\bibitem[{\citenamefont{Lenske}()}]{lenske_dfol}
\bibinfo{author}{\bibfnamefont{H.}~\bibnamefont{Lenske}},
  \emph{\bibinfo{title}{Dfol private comunication}}.

\bibitem[{\citenamefont{Franey and Love}(1985)}]{franey1985}
\bibinfo{author}{\bibfnamefont{M.~A.} \bibnamefont{Franey}} \bibnamefont{and}
  \bibinfo{author}{\bibfnamefont{W.~G.} \bibnamefont{Love}},
  \bibinfo{journal}{Phys. Rev.} \textbf{\bibinfo{volume}{C31}},
  \bibinfo{pages}{488} (\bibinfo{year}{1985}).

\bibitem[{\citenamefont{Goto}(2004)}]{rifuggiato2004}
\bibinfo{editor}{\bibfnamefont{A.}~\bibnamefont{Goto}}, ed.,
  \emph{\bibinfo{title}{{Cyclotrons and their applications. Proceedings, 17th
  International Conference, Cyclotrons 2004}}} (\bibinfo{publisher}{Part.
  Accel. Soc. Japan}, \bibinfo{address}{Tokyo, Japan}, \bibinfo{year}{2004}),
  \urlprefix\url{http://epaper.kek.jp/c04/index.html}.

\bibitem[{\citenamefont{Cappuzzello
  et~al.}(2016{\natexlab{b}})\citenamefont{Cappuzzello, Agodi, Carbone, and
  Cavallaro}}]{magnex_review}
\bibinfo{author}{\bibfnamefont{F.}~\bibnamefont{Cappuzzello}},
  \bibinfo{author}{\bibfnamefont{C.}~\bibnamefont{Agodi}},
  \bibinfo{author}{\bibfnamefont{D.}~\bibnamefont{Carbone}}, \bibnamefont{and}
  \bibinfo{author}{\bibfnamefont{M.}~\bibnamefont{Cavallaro}},
  \bibinfo{journal}{Eur. Phys. J.} \textbf{\bibinfo{volume}{A52}},
  \bibinfo{pages}{167} (\bibinfo{year}{2016}{\natexlab{b}}),
  \eprint{1606.06731}.

\bibitem[{\citenamefont{Cavallaro et~al.}(2012)\citenamefont{Cavallaro,
  Cappuzzello, Carbone, Cunsolo, Foti, Khouaja, Rodrigues, Winfield, and
  Bond{\`i}}}]{cavallaro2012}
\bibinfo{author}{\bibfnamefont{M.}~\bibnamefont{Cavallaro}},
  \bibinfo{author}{\bibfnamefont{F.}~\bibnamefont{Cappuzzello}},
  \bibinfo{author}{\bibfnamefont{D.}~\bibnamefont{Carbone}},
  \bibinfo{author}{\bibfnamefont{A.}~\bibnamefont{Cunsolo}},
  \bibinfo{author}{\bibfnamefont{A.}~\bibnamefont{Foti}},
  \bibinfo{author}{\bibfnamefont{A.}~\bibnamefont{Khouaja}},
  \bibinfo{author}{\bibfnamefont{M.~R.~D.} \bibnamefont{Rodrigues}},
  \bibinfo{author}{\bibfnamefont{J.~S.} \bibnamefont{Winfield}},
  \bibnamefont{and}
  \bibinfo{author}{\bibfnamefont{M.}~\bibnamefont{Bond{\`i}}},
  \bibinfo{journal}{The European Physical Journal A}
  \textbf{\bibinfo{volume}{48}}, \bibinfo{pages}{59} (\bibinfo{year}{2012}),
  ISSN \bibinfo{issn}{1434-601X},
  \urlprefix\url{https://doi.org/10.1140/epja/i2012-12059-8}.

\bibitem[{\citenamefont{Cappuzzello et~al.}(2010)\citenamefont{Cappuzzello,
  Cavallaro, Cunsolo, Foti, Carbone, Orrigo, and Rodrigues}}]{cappuzzello2010}
\bibinfo{author}{\bibfnamefont{F.}~\bibnamefont{Cappuzzello}},
  \bibinfo{author}{\bibfnamefont{M.}~\bibnamefont{Cavallaro}},
  \bibinfo{author}{\bibfnamefont{A.}~\bibnamefont{Cunsolo}},
  \bibinfo{author}{\bibfnamefont{A.}~\bibnamefont{Foti}},
  \bibinfo{author}{\bibfnamefont{D.}~\bibnamefont{Carbone}},
  \bibinfo{author}{\bibfnamefont{S.~E.~A.} \bibnamefont{Orrigo}},
  \bibnamefont{and} \bibinfo{author}{\bibfnamefont{M.~R.~D.}
  \bibnamefont{Rodrigues}}, \bibinfo{journal}{Nucl. Instrum. Meth.}
  \textbf{\bibinfo{volume}{A621}}, \bibinfo{pages}{419} (\bibinfo{year}{2010}).

\bibitem[{\citenamefont{Cappuzzello et~al.}(2011)\citenamefont{Cappuzzello,
  Carbone, and Cavallaro}}]{cappuzzello2011}
\bibinfo{author}{\bibfnamefont{F.}~\bibnamefont{Cappuzzello}},
  \bibinfo{author}{\bibfnamefont{D.}~\bibnamefont{Carbone}}, \bibnamefont{and}
  \bibinfo{author}{\bibfnamefont{M.}~\bibnamefont{Cavallaro}},
  \bibinfo{journal}{Nucl. Instrum. Methods Phys. Res. A}
  \textbf{\bibinfo{volume}{638}}, \bibinfo{pages}{74 } (\bibinfo{year}{2011}),
  ISSN \bibinfo{issn}{0168-9002},
  \urlprefix\url{http://www.sciencedirect.com/science/article/pii/S0168900211003585}.

\bibitem[{\citenamefont{Calabrese et~al.}(2018)}]{calabrese2018}
\bibinfo{author}{\bibfnamefont{S.}~\bibnamefont{Calabrese}}
  \bibnamefont{et~al.}, \bibinfo{journal}{Acta Phys. Polon.}
  \textbf{\bibinfo{volume}{B49}}, \bibinfo{pages}{275} (\bibinfo{year}{2018}).

\bibitem[{\citenamefont{Cavallaro et~al.}(2019)}]{cavallaro2019}
\bibinfo{author}{\bibfnamefont{M.}~\bibnamefont{Cavallaro}}
  \bibnamefont{et~al.}, \bibinfo{journal}{Res. in Phys.}
  \textbf{\bibinfo{volume}{13}}, \bibinfo{pages}{102191}
  (\bibinfo{year}{2019}), ISSN \bibinfo{issn}{2211-3797},
  \urlprefix\url{http://www.sciencedirect.com/science/article/pii/S2211379719302517}.

\bibitem[{\citenamefont{Cavallaro et~al.}(2011)\citenamefont{Cavallaro,
  Cappuzzello, Carbone, Cunsolo, Foti, and Linares}}]{cavallaro2011}
\bibinfo{author}{\bibfnamefont{M.}~\bibnamefont{Cavallaro}},
  \bibinfo{author}{\bibfnamefont{F.}~\bibnamefont{Cappuzzello}},
  \bibinfo{author}{\bibfnamefont{D.}~\bibnamefont{Carbone}},
  \bibinfo{author}{\bibfnamefont{A.}~\bibnamefont{Cunsolo}},
  \bibinfo{author}{\bibfnamefont{A.}~\bibnamefont{Foti}}, \bibnamefont{and}
  \bibinfo{author}{\bibfnamefont{R.}~\bibnamefont{Linares}},
  \bibinfo{journal}{Nucl. Instrum. Meth. A} \textbf{\bibinfo{volume}{637}},
  \bibinfo{pages}{77} (\bibinfo{year}{2011}).

\bibitem[{\citenamefont{Singh}(1995)}]{singh1995}
\bibinfo{author}{\bibfnamefont{B.}~\bibnamefont{Singh}},
  \bibinfo{journal}{Nuclear Data Sheets} \textbf{\bibinfo{volume}{74}},
  \bibinfo{pages}{63 } (\bibinfo{year}{1995}), ISSN \bibinfo{issn}{0090-3752},
  \urlprefix\url{http://www.sciencedirect.com/science/article/pii/S0090375285710058}.

\bibitem[{\citenamefont{Thompson}(1988)}]{thompson1988}
\bibinfo{author}{\bibfnamefont{I.~J.} \bibnamefont{Thompson}},
  \bibinfo{journal}{Comput. Phys. Rep.} \textbf{\bibinfo{volume}{7}},
  \bibinfo{pages}{167 } (\bibinfo{year}{1988}), ISSN \bibinfo{issn}{0167-7977},
  \urlprefix\url{http://www.sciencedirect.com/science/article/pii/0167797788900056}.

\bibitem[{\citenamefont{Cappuzzello et~al.}(2004)\citenamefont{Cappuzzello,
  Lenske, Cunsolo, Beaumel, Fortier, Foti, Lazzaro, Nociforo, Orrigo, and
  Winfield}}]{cappuzzello2004}
\bibinfo{author}{\bibfnamefont{F.}~\bibnamefont{Cappuzzello}},
  \bibinfo{author}{\bibfnamefont{H.}~\bibnamefont{Lenske}},
  \bibinfo{author}{\bibfnamefont{A.}~\bibnamefont{Cunsolo}},
  \bibinfo{author}{\bibfnamefont{D.}~\bibnamefont{Beaumel}},
  \bibinfo{author}{\bibfnamefont{S.}~\bibnamefont{Fortier}},
  \bibinfo{author}{\bibfnamefont{A.}~\bibnamefont{Foti}},
  \bibinfo{author}{\bibfnamefont{A.}~\bibnamefont{Lazzaro}},
  \bibinfo{author}{\bibfnamefont{C.}~\bibnamefont{Nociforo}},
  \bibinfo{author}{\bibfnamefont{S.~E.~A.} \bibnamefont{Orrigo}},
  \bibnamefont{and} \bibinfo{author}{\bibfnamefont{J.~S.}
  \bibnamefont{Winfield}}, \bibinfo{journal}{Nucl. Phys.}
  \textbf{\bibinfo{volume}{A739}}, \bibinfo{pages}{30} (\bibinfo{year}{2004}).

\bibitem[{\citenamefont{Ermamatov et~al.}(2016)\citenamefont{Ermamatov,
  Cappuzzello, Lubian, Cubero, Agodi, Carbone, Cavallaro, Ferreira, Foti
  et~al.}}]{ermamatov2016}
\bibinfo{author}{\bibfnamefont{M.}~\bibnamefont{Ermamatov}},
  \bibinfo{author}{\bibfnamefont{F.}~\bibnamefont{Cappuzzello}},
  \bibinfo{author}{\bibfnamefont{J.}~\bibnamefont{Lubian}},
  \bibinfo{author}{\bibfnamefont{M.}~\bibnamefont{Cubero}},
  \bibinfo{author}{\bibfnamefont{C.}~\bibnamefont{Agodi}},
  \bibinfo{author}{\bibfnamefont{D.}~\bibnamefont{Carbone}},
  \bibinfo{author}{\bibfnamefont{M.}~\bibnamefont{Cavallaro}},
  \bibinfo{author}{\bibfnamefont{J.}~\bibnamefont{Ferreira}},
  \bibinfo{author}{\bibfnamefont{A.}~\bibnamefont{Foti}}, \bibnamefont{et~al.},
  \bibinfo{journal}{Phys. Rev.} \textbf{\bibinfo{volume}{C94}},
  \bibinfo{pages}{024610} (\bibinfo{year}{2016}).

\bibitem[{\citenamefont{Carbone et~al.}(2017)}]{carbone2017}
\bibinfo{author}{\bibfnamefont{D.}~\bibnamefont{Carbone}} \bibnamefont{et~al.},
  \bibinfo{journal}{Phys. Rev.} \textbf{\bibinfo{volume}{C95}},
  \bibinfo{pages}{034603} (\bibinfo{year}{2017}).

\bibitem[{\citenamefont{Ermamatov et~al.}(2017)}]{ermamatov2017}
\bibinfo{author}{\bibfnamefont{M.~J.} \bibnamefont{Ermamatov}}
  \bibnamefont{et~al.}, \bibinfo{journal}{Phys. Rev.}
  \textbf{\bibinfo{volume}{C96}}, \bibinfo{pages}{044603}
  (\bibinfo{year}{2017}).

\bibitem[{\citenamefont{Paes et~al.}(2017)}]{paes2017}
\bibinfo{author}{\bibfnamefont{B.}~\bibnamefont{Paes}} \bibnamefont{et~al.},
  \bibinfo{journal}{Phys. Rev.} \textbf{\bibinfo{volume}{C96}},
  \bibinfo{pages}{044612} (\bibinfo{year}{2017}).

\bibitem[{\citenamefont{Pritychenko et~al.}(2016)\citenamefont{Pritychenko,
  Birch, Singh, and Horoi}}]{prytichenko_data_table}
\bibinfo{author}{\bibfnamefont{B.}~\bibnamefont{Pritychenko}},
  \bibinfo{author}{\bibfnamefont{M.}~\bibnamefont{Birch}},
  \bibinfo{author}{\bibfnamefont{B.}~\bibnamefont{Singh}}, \bibnamefont{and}
  \bibinfo{author}{\bibfnamefont{M.}~\bibnamefont{Horoi}},
  \bibinfo{journal}{Atom. Data Nucl. Data Tabl.}
  \textbf{\bibinfo{volume}{107}}, \bibinfo{pages}{1} (\bibinfo{year}{2016}),
  \eprint{1312.5975}.

\bibitem[{\citenamefont{Fonseca et~al.}(2019)\citenamefont{Fonseca, Linares,
  Zagatto, Cappuzzello, Carbone, Cavallaro, Agodi, Lubian, and
  Oliveira}}]{fonseca2019}
\bibinfo{author}{\bibfnamefont{L.~M.} \bibnamefont{Fonseca}},
  \bibinfo{author}{\bibfnamefont{R.}~\bibnamefont{Linares}},
  \bibinfo{author}{\bibfnamefont{V.~A.~B.} \bibnamefont{Zagatto}},
  \bibinfo{author}{\bibfnamefont{F.}~\bibnamefont{Cappuzzello}},
  \bibinfo{author}{\bibfnamefont{D.}~\bibnamefont{Carbone}},
  \bibinfo{author}{\bibfnamefont{M.}~\bibnamefont{Cavallaro}},
  \bibinfo{author}{\bibfnamefont{C.}~\bibnamefont{Agodi}},
  \bibinfo{author}{\bibfnamefont{J.}~\bibnamefont{Lubian}}, \bibnamefont{and}
  \bibinfo{author}{\bibfnamefont{J.~R.~B.} \bibnamefont{Oliveira}},
  \bibinfo{journal}{Phys. Rev. C} \textbf{\bibinfo{volume}{100}},
  \bibinfo{pages}{014604} (\bibinfo{year}{2019}),
  \urlprefix\url{https://link.aps.org/doi/10.1103/PhysRevC.100.014604}.

\bibitem[{\citenamefont{Crema et~al.}(2011)\citenamefont{Crema, Otomar,
  Sim\~oes, Barioni, Monteiro, Ono, Shorto, Lubian, and Gomes}}]{crema2011}
\bibinfo{author}{\bibfnamefont{E.}~\bibnamefont{Crema}},
  \bibinfo{author}{\bibfnamefont{D.~R.} \bibnamefont{Otomar}},
  \bibinfo{author}{\bibfnamefont{R.~F.} \bibnamefont{Sim\~oes}},
  \bibinfo{author}{\bibfnamefont{A.}~\bibnamefont{Barioni}},
  \bibinfo{author}{\bibfnamefont{D.~S.} \bibnamefont{Monteiro}},
  \bibinfo{author}{\bibfnamefont{L.~K.} \bibnamefont{Ono}},
  \bibinfo{author}{\bibfnamefont{J.~M.~B.} \bibnamefont{Shorto}},
  \bibinfo{author}{\bibfnamefont{J.}~\bibnamefont{Lubian}}, \bibnamefont{and}
  \bibinfo{author}{\bibfnamefont{P.~R.~S.} \bibnamefont{Gomes}},
  \bibinfo{journal}{Phys. Rev. C} \textbf{\bibinfo{volume}{84}},
  \bibinfo{pages}{024601} (\bibinfo{year}{2011}),
  \urlprefix\url{https://link.aps.org/doi/10.1103/PhysRevC.84.024601}.

\bibitem[{\citenamefont{Crema et~al.}(2018)\citenamefont{Crema, Zagatto,
  Shorto, Paes, Lubian, Sim\~oes, Monteiro, Huiza, Added et~al.}}]{crema2018}
\bibinfo{author}{\bibfnamefont{E.}~\bibnamefont{Crema}},
  \bibinfo{author}{\bibfnamefont{V.~A.~B.} \bibnamefont{Zagatto}},
  \bibinfo{author}{\bibfnamefont{J.~M.~B.} \bibnamefont{Shorto}},
  \bibinfo{author}{\bibfnamefont{B.}~\bibnamefont{Paes}},
  \bibinfo{author}{\bibfnamefont{J.}~\bibnamefont{Lubian}},
  \bibinfo{author}{\bibfnamefont{R.~F.} \bibnamefont{Sim\~oes}},
  \bibinfo{author}{\bibfnamefont{D.~S.} \bibnamefont{Monteiro}},
  \bibinfo{author}{\bibfnamefont{J.~F.~P.} \bibnamefont{Huiza}},
  \bibinfo{author}{\bibfnamefont{N.}~\bibnamefont{Added}},
  \bibnamefont{et~al.}, \bibinfo{journal}{Phys. Rev. C}
  \textbf{\bibinfo{volume}{98}}, \bibinfo{pages}{044614}
  (\bibinfo{year}{2018}),
  \urlprefix\url{https://link.aps.org/doi/10.1103/PhysRevC.98.044614}.

\end{thebibliography}

\end{document}